\documentclass[12pt, a4paper]{article}
\usepackage{cite}
\usepackage{amsmath,amssymb}
\input{colordvi.tex}
\usepackage{color}
\usepackage{graphicx}
\usepackage{comment}
\usepackage[colorlinks=true,linkcolor=red,urlcolor=blue,citecolor=blue]{hyperref}

\begin{document}

\begin{titlepage}

\begin{center}

\hfill DESY 19-050\\
\hfill KEK-TH-2114\\
\hfill UT-19-04\\

\vskip .75in

{\Large \bf 
Production of Purely Gravitational Dark Matter: \\ \vspace{2mm}  The Case of Fermion and Vector Boson
}

\vskip .75in

{\large
Yohei Ema$^{(a,b)}$, Kazunori Nakayama$^{(c,d)}$ and Yong Tang$^{(c)}$
}

\vskip 0.25in

$^{(a)}${\em DESY, Notkestra{\ss}e 85, D-22607 Hamburg, Germany}\\[.3em]
$^{(b)}${\em KEK Theory Center, Tsukuba 305-0801, Japan}\\[.3em]
$^{(c)}${\em Department of Physics, Faculty of Science,\\
The University of Tokyo,  Bunkyo-ku, Tokyo 113-0033, Japan}\\[.3em]
$^{(d)}${\em Kavli IPMU (WPI), The University of Tokyo,  Kashiwa, Chiba 277-8583, Japan}

\end{center}
\vskip .5in

\begin{abstract}

We consider the simplest possibility for a model of particle dark matter in which dark matter has only gravitational interaction with the standard model sector. Even in such a case, it is known that the gravitational particle production in an expanding universe may lead to a correct relic abundance depending on the inflation scale and the mass of dark matter particle. We provide a comprehensive and systematic analysis of the gravitational particle production of fermionic and vectorial dark matter, and emphasize that particles which are much heavier than the Hubble parameter but lighter than inflaton can also be produced abundantly.

\end{abstract}

\end{titlepage}


\renewcommand{\thepage}{\arabic{page}}
\setcounter{page}{1}
\renewcommand{\thefootnote}{\#\arabic{footnote}}
\setcounter{footnote}{0}

\newpage

\tableofcontents

\section{Introduction}
\label{sec:Intro}

Gravitational particle production is a universal phenomenon which happens in the expanding universe~\cite{Parker:1969au,Birrell:1982ix}. If a field is not conformal, it feels the time-dependent background evolution of the metric and results in finite number density even if we start with the adiabatic vacuum with no particle excitation initially. 
Gravitational particle production in inflationary universe was discussed in Refs.~\cite{Ford:1986sy,Chung:1998zb,Chung:2001cb}, where it was shown that the particle production happens during the transition from inflationary (or de Sitter) era to the radiation-dominated or matter-dominated universe. Depending on the mass of the particle, these gravitationally produced particles can be dominant component of dark matter (DM). A typical number density of produced particles during the transition is given by $n \sim H_{\rm inf}^3$ where $H_{\rm inf}$ denotes the Hubble parameter during inflation for a scalar field with a minimal coupling to gravity if $m\lesssim H_{\rm inf}$ where $m$ denotes the scalar field mass.\footnote{
	Precisely, $H_{\rm inf}$ should be regarded as the Hubble scale at the end of inflation. In this paper we take the Hubble parameter $H$ to be constant during inflation, which is a good approximation for our purpose.
}
Note also that a light scalar field with $m\lesssim H_{\rm inf}$ gravitationally develops condensation during inflation, whose subsequent oscillation can account for the DM in the present universe (see e.g. Refs.~\cite{Markkanen:2018gcw,Alonso-Alvarez:2018tus} for recent works on general spectator fields).

Recently it was argued that the gravitational particle production also happens during the reheating era in which the inflaton oscillates coherently, due to the fact that the cosmic scale factor $a(t)$ contains a small but rapidly oscillating part~\cite{Ema:2015dka,Ema:2016hlw}. Since the time scale of the coherent oscillation is given by $\sim m_{\rm inf}^{-1}$, where $m_{\rm inf}$ denotes the inflaton mass around its potential minimum, particles with mass smaller than $m_{\rm inf}$ are likely to be produced, along the line of preheating in models where the inflaton directly couples to other fields~\cite{Dolgov:1989us,Traschen:1990sw,Shtanov:1994ce,Kofman:1997yn}. The number density produced in this way is also given by $n\sim H_{\rm inf}^3$ for a minimally coupled scalar. It is comparable to the contribution discussed above, but now the particle mass can be as large as the inflaton mass $m_{\rm inf}$. Since many inflation models have $m_{\rm inf}\gg H_{\rm inf}$, it opens up a possibility that the super-Hubble mass particles are efficiently produced and become a dominant component of DM. This scenario was further studied in detail in Refs.~\cite{Ema:2018ucl,Chung:2018ayg}. See also Refs.~\cite{Hashiba:2018tbu,Li:2019ves} for some related works.\footnote{
	Gravitationally interacting particles are also produced by scatterings of the Standard Model (SM) particles in thermal bath through the graviton exchange~\cite{Garny:2015sjg,Tang:2016vch,Tang:2017hvq,Garny:2017kha}. 
}
This class of scenario is interesting since we do not need any particular interaction of DM with SM particles to account for the present DM abundance. In this sense, it may be the simplest model of DM.

In this paper we consider production of a massive fermion and vector boson which only  have gravitational interaction.
A fermion or vector boson may be more suitable as a candidate of purely gravitational DM since their interactions with the SM particles can be naturally forbidden at the renormalizable level. For a singlet fermion $\psi$, the only possible renormalizable interaction in the Lagrangian is $\mathcal L \sim y_i \psi (L_i \Phi) + {\rm h.c.}$ where $L_i$ $(i=1,2,3)$ denotes the lepton doublet and $\Phi$ is the SM Higgs boson with $y_i$ being coupling constants. 
This coupling is forbidden by, e.g., assuming that the SM fermions are charged under $B-L$ gauge symmetry but $\psi$ is a singlet.
Especially, $\psi$ is absolutely stable if the $Z_2$ subgroup of U(1) $B-L$ symmetry remains unbroken.
For an Abelian hidden vector boson $A_\mu$, the only possible renormalizable interaction is the kinetic mixing with hypercharge photon, which may also be forbidden by imposing discrete symmetry. For a scalar field $\phi$, on the other hand, one can always introduce a Higgs-portal coupling $\mathcal L \sim - \lambda \phi^2 |\Phi|^2$ which drives $\phi$ into thermal bath unless the coupling constant $\lambda$ is very small. Once thermalized, the DM abundance is determined by the standard freezeout scenario~\cite{Silveira:1985rk,McDonald:1993ex}.
Thus fermion or vector boson can be more likely to be purely gravitational DM.

The gravitational production of fermion or vector boson is qualitatively different from the case of scalar field with minimal coupling to gravity.
This is because the fermion and (transverse) vector boson are conformal in the massless limit. One can choose a canonical basis such that the dependence on the cosmic scale factor completely disappears. Thus there is no particle production in the massless limit, which is similar to the case of a scalar field with conformal coupling. The case of massive vector boson is a bit complicated since the longitudinal component behaves rather like a minimally coupled scalar field and requires a careful investigation.

In Sec.~\ref{sec:fermion} we study the gravitational fermion production. 
In Sec.~\ref{sec:vector} we investigate the gravitational production of massive vector boson in detail.
In particular, we carefully distinguish the transverse and longitudinal mode and discuss how they behave under the time-dependent background.
We conclude in Sec.~\ref{sec:conc}.

\section{Fermion production}
\label{sec:fermion}

\subsection{Fermion action in the FRW Universe}

Let us consider an action of free Majorana fermion $\psi$, which satisfies the Majorana condition $\psi=-C^{-1}\overline\psi^T$ with $C$ denoting the charge conjugation matrix,
\begin{align}
	S = \int d^4x \,e \left[-\frac{1}{2}\overline\psi\left(e^\mu_a \gamma^a D_\mu - m\right)\psi \right],
\end{align}
where $e^\mu_a$ denotes the vierbein with $a,b,\dots$ and $\mu,\nu,\dots$ represent local Lorentz and general coordinate indices respectively,
and $e\equiv \det(e^\mu_a)=\sqrt{-g}$.
The covariant derivative is given by
\begin{align}
	D_\mu = \partial_\mu + \frac{1}{4} \omega_\mu^{ab}\gamma_{[a}\gamma_{b]},
\end{align}
where the spin connection is defined as
\begin{align}
	\omega_\mu^{ab} = 2e^{\nu [a}\partial_{[\mu}{e_{\nu]}}^{b]} - e^{\nu[a}e^{b]\sigma}e_{\mu c}\partial_\nu e_{\sigma}^c.
\end{align}
In the Friedmann-Robertson-Walker (FRW) background, 
\begin{align}
	ds^2 = -dt^2+a^2(t)\delta_{ij}dx^idx^j = a^2(\tau) \left( -d\tau^2 + \delta_{ij}dx^idx^j \right),
\end{align}
the only non-zero components are
\begin{align}
	{\omega_i}^{j0}=\delta_i^j \mathcal H,
\end{align}
where $\mathcal H = a'/a$ is the conformal Hubble parameter, which is related to the Hubble parameter $H = \dot a/a$ through $\mathcal H = aH$. Here and in what follows, the prime (dot) denotes the derivative with respect to conformal time $\tau$ (physical time $t$).
It is convenient to perform the rescaling as $\widetilde \psi \equiv a^{3/2} \psi$ so that the action becomes
\begin{align}
	S = \int d\tau d^3x \left[-\frac{1}{2}\overline{\widetilde\psi}\left(\delta^\mu_a \gamma^a \partial_\mu - a m\right)\widetilde\psi \right].
\end{align}
It is seen that the rescaled field has a canonical kinetic term and the action is independent of the scale factor $a$ in the massless limit $m\to 0$.  In other words, a fermion is conformal in the massless limit. Therefore, the rate of gravitational particle production is suppressed by the fermion mass $m_\psi$.

It is convenient to work with the Fourier mode since we are interested in the free fermion:
\begin{align}
	\widetilde\psi(\vec x,\tau) = \int \frac{d^3k}{(2\pi)^3} \psi_{\vec k}(\tau) e^{i \vec k\cdot \vec x}.
\end{align}
The equation of motion for the mode function is given by
\begin{align}
	\left( \partial_\tau \gamma^0 - i \vec k\cdot \vec \gamma - am \right) \psi_{\vec k} (\tau) = 0.
\end{align}
Now let us expand the mode function as
\begin{align}
	 \psi_{\vec k} (\tau) = \sum_{h=\pm} \left[ u_{\vec k, h}(\tau)  b_{\vec k, h} + v_{\vec k,h}(\tau)  b^\dagger_{-\vec k,h} \right],
\end{align}
where $v_{\vec k,h} = -C^{-1} \overline{u}^T_{-\vec k,h}$ and $h$ denotes the helicity degree of freedom. The normalization condition is taken as follows:
\begin{align}
	u^\dagger_{\vec k,h}(\tau) u_{\vec k,h'}(\tau) =v^\dagger_{\vec k,h}(\tau) v_{\vec k,h'}(\tau) = \delta_{hh'},~~~~~~
	u^\dagger_{\vec k,h}(\tau) v_{\vec k,h'}(\tau) = 0.
\end{align}
The creation and annihilation operators satisfy the following anti-commutation relation:
\begin{align}
	\left\{ b_{\vec k,h},b^\dagger_{\vec k',h'} \right\} = \left(2\pi\right)^3\delta(\vec k - \vec k') \delta_{hh'},~~~~~~
	\left\{ b_{\vec k,h},b_{\vec k',h'} \right\} = \left\{ b^\dagger_{\vec k,h},b^\dagger_{\vec k',h'} \right\} = 0,
\end{align}
so that the original field satisfies the anti-commutation relation $\left\{ \widetilde\psi(\tau,\vec x),\widetilde\psi^\dagger(\tau,\vec y) \right\} = \delta(\vec x-\vec y)$. Let us write the mode function as
\begin{align}
	u_{\vec k,h}(\tau) =\begin{pmatrix}
		u_{\vec k,h}^+(\tau) \\
		u_{\vec k,h}^-(\tau)
	\end{pmatrix} 
	\otimes \xi_{\vec k,h},
\end{align}
where $ \xi_{\vec k,h}$ denotes the eigenvector of the helicity, which satisfies $(\vec \sigma\cdot \hat{\vec k})  \xi_{\vec k,h} = h  \xi_{\vec k,h}$ with $\hat k \equiv \vec k /|\vec k|$ and $h=\pm 1$. Taking $\vec k$ to be the $z$-direction, we have $\xi_{\vec k,+}=(1,0)^T$ and $\xi_{\vec k,-}=(0,1)^T$. Adopting the Dirac representation for the gamma matrices, the equation of motion becomes 
\begin{align}
	i\partial_\tau u_{\vec k,h}^{\pm}(\tau)+ hk u_{\vec k,h}^{\mp}(\tau) \mp am u_{\vec k,h}^{\pm}(\tau) = 0,
\end{align}
which may be cast into the second order form,
\begin{align}
	\partial_\tau^2 u_{\vec k,h}^{\pm}(\tau)+\left[ \omega_k^2(\tau)\pm i (am)' \right] u_{\vec k,h}^{\pm}(\tau) =0,~~~~~~\omega_k^2(\tau)\equiv k^2+a^2m^2.   \label{eom_u}
\end{align}
Note that $u_{\vec k,h}^{+}$ and $u_{\vec k,h}^{-}$ are not independent of each other and the normalization condition implies
\begin{align}
	\left| u_{\vec k,h}^{+}(\tau) \right|^2+\left| u_{\vec k,h}^{-}(\tau) \right|^2 = 1.  \label{normalization}
\end{align}

During the de Sitter phase in which the Hubble parameter is given by $H=H_{\rm inf}={\rm const}$, noting $(am)'=a^2H_{\rm inf}m$ and $\tau=-1/(aH_{\rm inf})$, we find an exact solution to Eq.~(\ref{eom_u}) as
\begin{align}
	u_{\vec k,h}^{\pm}(\tau) = i \epsilon\sqrt{\frac{-\pi k\tau}{4}}e^{\pm \frac{\pi m}{2 H_{\rm inf}}} H_{\nu_{\pm}}^{(1)}(-k\tau),~~~~~~
	\nu_{\pm} \equiv \frac{1}{2}\mp \frac{im}{H_{\rm inf}},
\end{align}
where $\epsilon=1$ for $u_{\vec k,h}^+$ and $\epsilon = -h$ for $u_{\vec k,h}^-$, and $H_{\nu}^{(1)}(x)$ is the Hankel function of the first kind with order $\nu$. In the far past $(k\tau\to -\infty)$, or the short wavelength limit, it approaches to the same mode function as the Minkowski vacuum:
\begin{align}
	u_{\vec k,h}^+ \to \sqrt{\frac{\omega_k + am}{2\omega_k}} e^{-i\int \omega_k d\tau},~~~~~~
	u_{\vec k,h}^- \to -h\sqrt{\frac{\omega_k - am}{2\omega_k}} e^{-i\int \omega_k d\tau}.  \label{u_vacuum}
\end{align}
Here $a(\tau) \to 0$ limit should be understood. It is also evident that there is no significant growth in the superhorizon limit $(k\tau \to 0)$ during inflation.\footnote{
Asymptotic form of the Hankel function in the short wavelength limit is
\begin{align}
	H_\nu^{(1)}(x) \to \sqrt{\frac{2}{\pi x}}e^{i(x-(2\nu+1)\pi/4)} ~~~{\rm for}~~~x\to +\infty.
\end{align}
In the long wavelength limit, the Hankel function becomes
\begin{align}
	H_\nu^{(1)}(x) \to -\frac{i}{\pi}\Gamma(\nu)\left(\frac{2}{x}\right)^{\nu}~~{\rm for}~~x\to 0~~{\rm and}~~{\rm Re(\nu)}>0.
\end{align}
}

\subsection{Fermion production}

Now let us estimate the gravitational fermion production. The fermion production in the rapidly oscillating background or the fermionic preheating was studied in Refs.~\cite{Greene:1998nh,Greene:2000ew,Peloso:2000hy,Asaka:2010kv}. The gravitational fermion production in the expanding universe was studied in Refs.~\cite{Lyth:1996yj,Kuzmin:1998kk,Chung:2011ck} and also the gravitino preheating was extensively studied in Refs.~\cite{Maroto:1999ch,Kallosh:1999jj,Giudice:1999am,Kallosh:2000ve,Nilles:2001ry,Nilles:2001fg,Ema:2016oxl}. Here we combine these knowledges to estimate the rate of gravitational fermion production, especially pointing out the contribution from the inflaton coherent oscillation.

The energy density of the fermion is given by
\begin{align}
	a^4(\tau)\rho_\psi(\tau) &= \frac{1}{2}\left< \widetilde\psi^\dagger i \partial_\tau \widetilde\psi \right> 
	= 2\int\frac{d^3k}{(2\pi)^3} \omega_k f_\psi(k,\tau) 
\end{align}
where the prefactor $2$ counts the two helicity modes and 
\begin{align}
	f_\psi(\vec k,\tau) &\equiv \frac{1}{2\omega_k}\left[
		am \left( \left|u^-_{\vec k,h}(\tau)\right|^2 -  \left|u^+_{\vec k,h}(\tau)\right|^2 \right) + 2h k \,{\rm Re}\left( u_{\vec k,h}^+(\tau)u_{\vec k,h}^{- *}(\tau) \right)
	\right] + \frac{1}{2}, \\
	&= \frac{1}{2\omega_k}\left[
		am + 2\,{\rm Im}\left( u_{\vec k,h}^{+*}(\tau)\partial_\tau u_{\vec k,h}^{+}(\tau) \right)
	\right] + \frac{1}{2}
\end{align}
denotes the occupation number or the phase space density. The last factor $+1/2$ cancels the negatively divergent energy density due to the fermionic zero-point fluctuations. One sees that $f_\psi=0$ for the Minkowski mode function (\ref{u_vacuum}). In the time-dependent background $(a' > 0)$ the mode function may be modified from this asymptotic form and hence we will obtain $f_\psi > 0$ that signals particle production.
The number density is also a useful quantity, which is then given by
\begin{align}
	a^3(\tau)n_\psi(\tau) =  2\int\frac{d^3k}{(2\pi)^3} f_\psi(k,\tau).
\end{align}

In order to estimate the particle production, we conveniently rewrite the mode function as
\begin{align}
	u_{\vec k, h}^+(\tau) = A_{k, h}(\tau)g_+ e^{-i\int^\tau \omega_k(\tau') d\tau'}
	+ B_{k, h}(\tau)g_- e^{i\int^\tau\omega_k(\tau') d\tau'},
\end{align}
where coefficients are assumed to satisfy
\begin{align}
	A_{k, h}'(\tau) = -\frac{g_-'}{g_+}e^{2i\int^\tau \omega_k(\tau') d\tau'} B_{k, h}(\tau),~~~~~~
	B_{k, h}'(\tau) = -\frac{g_+'}{g_-}e^{-2i\int^\tau \omega_k(\tau') d\tau'} A_{k, h}(\tau),
	\label{eq:diff_AB}
\end{align}
where $g_\pm \equiv \sqrt{ (\omega_k\pm am)/(2\omega_k) }$.
One can check that $u_{k,h}^+(\tau)$ satisfies the equation of motion (\ref{eom_u}). 
The other mode function is given by
\begin{align}
	u_{\vec k, h}^-(\tau) = -h \left[ A_{k,h}(\tau)g_- e^{-i\int^\tau \omega_k(\tau') d\tau'}
	- B_{k,h}(\tau)g_+ e^{i\int^\tau\omega_k(\tau') d\tau'}\right],
\end{align}
and hence the normalization condition (\ref{normalization}) implies $|A_{k,h}(\tau)|^2+|B_{k,h}(\tau)|^2=1$, which ensures that the phase space density cannot exceed unity as expected from the Pauli exclusion principle. 
The initial condition~\eqref{u_vacuum} is equivalent to $A_{k,h}(\tau\to -\infty) = 1$ and $B_{k,h}(\tau\to-\infty)= 0$.
Since the initial condition and the time evolution~\eqref{eq:diff_AB} 
do not explicitly depend on $h$, $A_{k,h}$ and $B_{k,h}$ are 
the same for both $h=\pm$. In what follows we omit the helicity subscript $h$ for this reason.
Deviation from these initial values $A_k = 1$ and $B_k = 0$ indicate particle production. With these definitions, we have
\begin{align}
	f_\psi(\vec k,\tau) = |B_k(\tau)|^2,~~~~~~
	B_k(\tau) \simeq \int^\tau d\tau' \frac{a^2 Hmk}{2\omega_k^2}e^{-2i\int^{\tau'}\omega_k(\tau'') d\tau''}.
\end{align}
Thus what we have to do is to calculate $B_k(\tau)$ under the background evolution of the cosmic scale factor $a(\tau)$. The calculation is almost parallel to the case of scalar field with conformal coupling. Thus we do not write the detail of the calculation below. Readers are referred to Ref.~\cite{Ema:2018ucl} for more detailed discussion to evaluate this integral.
There it was shown that there are two contributions to the particle production: one comes from the ``slow'' change of the background whose time scale is just parametrized by the Hubble expansion rate $H$ and the other comes from the ``fast'' change of the background related to the inflaton coherent oscillation whose time scale is characterized by the inflaton mass $m_\phi$~\cite{Ema:2015dka,Ema:2016hlw}.

For the ``slow'' contribution, the production rate is suppressed as $e^{-cm/H_{\rm inf}}$ with $c$ being order one numerical constant for $m\gtrsim H_{\rm inf}$. Thus we focus on the case $m \lesssim H_{\rm inf}$. In this case, the dominant production comes from the epoch around $k\sim a m$, i.e., the fermion becomes non-relativistic. The number density produced in this way is estimated as
\begin{align}
	n_\psi^{\rm (slow)}(\tau) \sim \mathcal A \,a^{-3}(\tau) \,k_c^3,~~~~~~
	\frac{k_c}{a_{\rm end}} \equiv m \left( \frac{H_{\rm inf}}{m} \right)^{\frac{2}{3(1+w)}},  \label{kc}
\end{align}
where $\mathcal A \sim \mathcal O(10^{-3})$ is a numerical coefficient and $k_c$ is defined as a momentum such that $k/a(\tau) = m$ at $H = m$, $a_{\rm end}$ denotes the scale factor at the end of inflation and $w$ denotes the equation of state parameter after inflation until the epoch $H=m$. Hereafter we take $w=0$ for simplicity, having in mind a scenario that the inflaton coherent oscillation behaves as non-relativistic matter and it finally decays into radiation at $H=H_{\rm R}$ with the reheating temperature $T_{\rm R} =(90/\pi^2 g_*)^{1/4} \sqrt{H_{\rm R} M_P}$.\footnote{
	We implicitly assumed that reheating is competed after $H=m$, i.e., $H_{\rm R} < m$. Otherwise, $k_c$ should be replaced by $k_c/a_{\rm end}=m(H_{\rm inf}/m)^{2/3} (m/H_{\rm R})^{1/6}$.
} This result is consistent with Ref.~\cite{Chung:2011ck}.
The ``fast'' contribution is given by
\begin{align}
	n_\psi^{\rm (fast)}(\tau) \sim \mathcal C H_{\rm inf}^3 \left( \frac{m}{m_{\rm inf}} \right)^2 \left( \frac{a_{\rm end}}{a(\tau)} \right)^3,
\end{align}
where $m_{\rm inf}$ denotes the inflaton mass and $\mathcal C \sim 10^{-3}-10^{-2}$ is a numerical constant. As emphasized in Refs.~\cite{Ema:2015dka,Ema:2016hlw,Ema:2018ucl,Chung:2018ayg}, this fast contribution is not suppressed even for $m\gg H_{\rm inf}$ but gets a suppression for $m\gtrsim m_{\rm inf}$ and the majority of inflation models predicts $m_{\rm inf} \gg H_{\rm inf}$. The final fermion abundance is roughly given by the sum of these two contributions. In any case, the fermion abundance goes to zero in the massless limit $m\to 0$ as expected, since a fermion is conformal in this limit.  
Combining them, the present fermion abundance in terms of its energy density divided by the entropy density $s$ is given by
\begin{align}
	\frac{\rho_\psi}{s} \simeq \frac{m H_{\rm inf} T_{\rm R}}{4 M_P^2}\left[ \mathcal C \left( \frac{m}{m_{\rm inf}} \right)^2
	+ \eta \frac{m}{H_{\rm inf}}\right],
\end{align}
where
\begin{align}
	\eta = \mathcal A \times \begin{cases}
		1  & {\rm for} ~~m\lesssim H_{\rm inf} \\
		(m/H_{\rm inf})^2 e^{-cm/H_{\rm inf}}   & {\rm for} ~~m\gtrsim H_{\rm inf}
	\end{cases}.  \label{eta}
\end{align}
For $H_{\rm inf}\ll m \lesssim m_{\rm inf}$, the fast contribution likely dominates and we have
\begin{align}
	\frac{\rho_\psi}{s} \simeq 4\times 10^{-10}{\rm GeV}\,\mathcal C 
	\left( \frac{m}{10^9\,{\rm GeV}} \right)\left( \frac{H_{\rm inf}}{10^9\,{\rm GeV}} \right)\left( \frac{T_{\rm R}}{10^{10}\,{\rm GeV}} \right)
	\left( \frac{m}{m_{\rm inf}} \right)^2.
\end{align}
For $m \lesssim H_{\rm inf}$, on the other hand, the slow contribution likely dominates and we have
\begin{align}
	\frac{\rho_\psi}{s} \simeq 4\times 10^{-10}{\rm GeV}\,\mathcal A 
	\left( \frac{m}{10^9\,{\rm GeV}} \right)^2 
	\left( \frac{T_{\rm R}}{10^{10}\,{\rm GeV}} \right).   \label{rhopsi_slow}
\end{align}
In this case, the abundance is independent of the inflationary scale $H_{\rm inf}$.\footnote{
	If $m < H_{\rm R}$, the expression (\ref{rhopsi_slow}) should be multiplied by the factor $(m/H_{\rm R})^{1/2}$ (hence becomes independent of the reheating temperature) due to the change of expression of $k_c$, as noticed in the previous footnote.
}
Comparing it with the present DM abundance $\rho_{\rm DM}/s \sim 4\times 10^{-10}\,{\rm GeV}$, it is possible that a pure singlet fermion having only the gravitational interaction becomes a dominant component of DM if its mass is relatively large.

\begin{figure}[t]
	\begin{center}
		\includegraphics[scale=0.35]{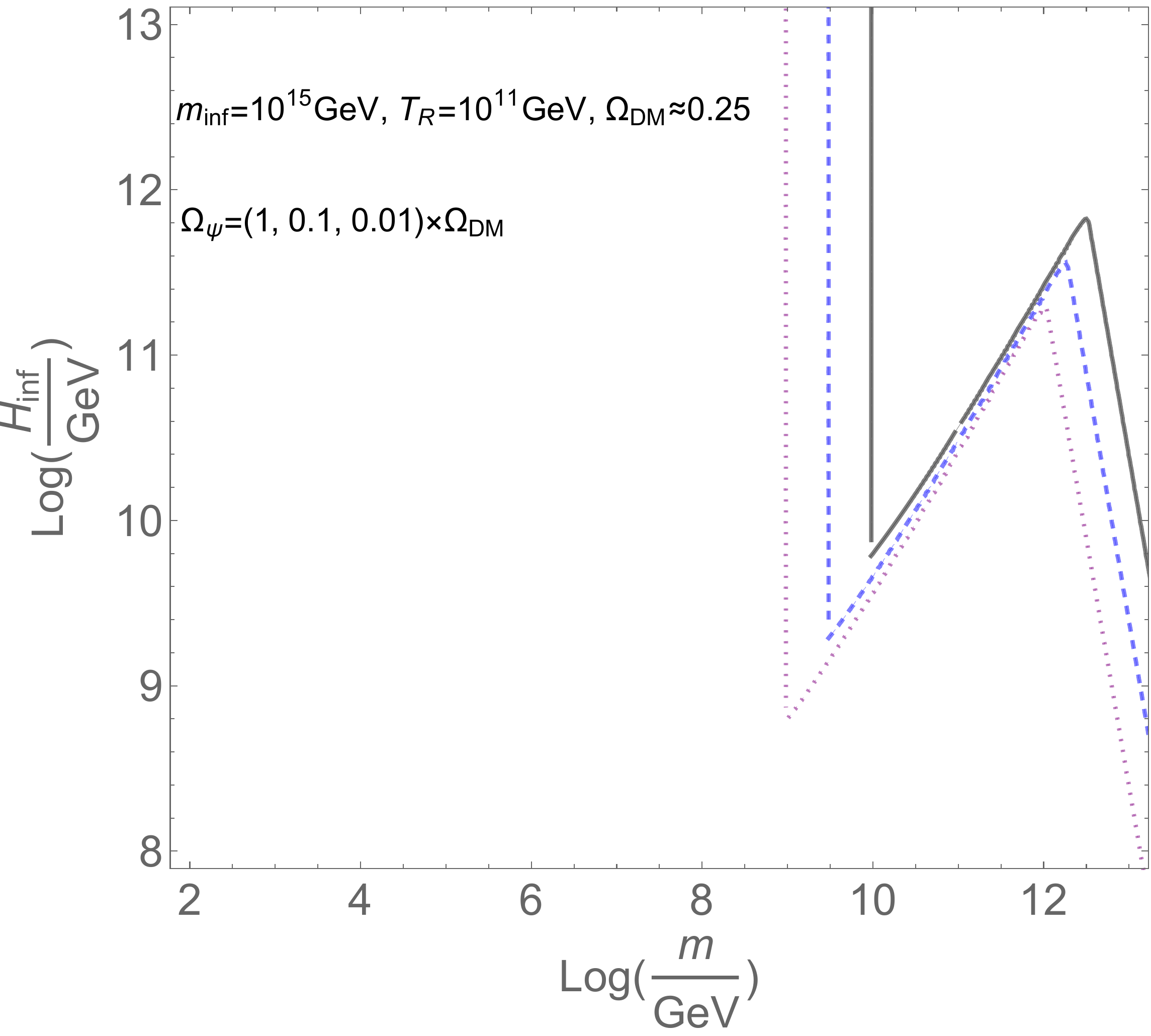}
		\includegraphics[scale=0.35]{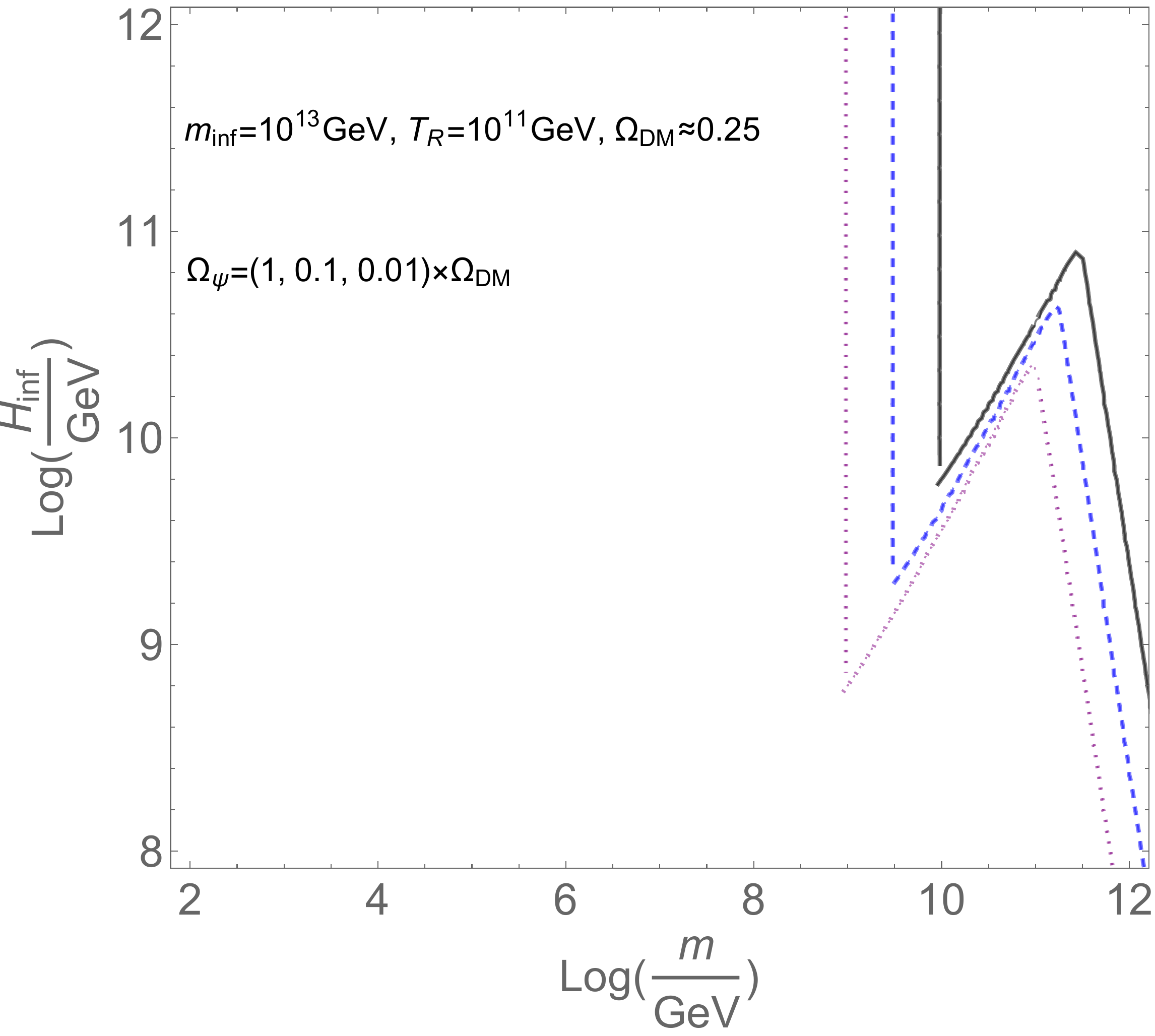}
	\end{center}
	\caption{Illustration  of the gravitationally produced fermion abundance with two sets of inflaton mass and reheating temperature, $m_{\textrm{inf}}=10^{15}\,$GeV and $T_{\rm R}=10^{11}\,$GeV (Left), $m_{\textrm{inf}}=10^{13}\,$GeV and $T_{\rm R}=10^{11}\,$GeV (Right). Three different curves (gray solid, blue dashed, and purple dotted) correspond to $\Omega_{\psi}=(1,0.1,0.01)\times\Omega_{\textrm{DM}}$. 
		\label{fig:fermion}	}
\end{figure}

In Fig.~\ref{fig:fermion} we illustrate with several contours of the fermion abundance on the plane of $(m,H_{\rm inf})$ with two choices of inflaton mass and reheating temperature, the left panel with $m_{\textrm{inf}}=10^{15}\,$GeV and $T_{\rm R}=10^{11}\,$GeV,　 and right panel with $m_{\textrm{inf}}=10^{13}\,$GeV and $T_{\rm R}=10^{11}\,$GeV. Three different contours (gray solid, blue dashed, and purple dotted) correspond to $\Omega_{\psi}=(1,0.1,0.01)\times\Omega_{\textrm{DM}}$, where $\Omega_{i}$ $(i=\psi,{\rm DM})$ is the density parameter defined by $\rho_{i} / \rho_{\rm crit}$ with $\rho_{\rm crit}$ being the critical energy density of the present universe. Evidently, wide parameter space exists for the correct DM relic density. 
Note that we should also include the contribution from thermal production by gravitational annihilation of SM particles in the thermal bath~\cite{Garny:2015sjg,Tang:2016vch,Tang:2017hvq,Garny:2017kha}. See Appendix~\ref{sec:thermalp} for details. In the parameter space we have shown, however, contributions from thermal production is negligible.

\section{Vector boson production}
\label{sec:vector}

\subsection{Vector boson action in the FRW Universe}

Let us consider an action of massive vector boson,
\begin{align}
	S = \int d^4x \sqrt{-g} \left[-\frac{1}{4}g^{\mu\rho} g^{\nu\sigma}F_{\mu\nu}F_{\rho\sigma} - \frac{1}{2}m^2 g^{\mu\nu}A_\mu A_\nu \right],
	\label{massive_vector}
\end{align}
where $F_{\mu\nu} = \partial_\mu A_\nu-\partial_\nu A_\mu$. In the FRW background, this action is rewritten as
\begin{align}
	S = \int d\tau d^3x \left[-\frac{1}{4}\eta^{\mu\rho} \eta^{\nu\sigma}F_{\mu\nu}F_{\rho\sigma} - \frac{1}{2}a^2 m^2 \eta^{\mu\nu}A_\mu A_\nu \right].
\end{align}
One can impose a $Z_2$ symmetry under which only $A_\mu$ changes its sign to forbid the kinetic mixing with the standard model hypercharge photon. Then $A_\mu$ is stable and a candidate of DM. See Refs.~\cite{Lebedev:2011iq,Farzan:2012hh} for concrete model buildings.

The vector boson mass can be regarded as a result of the Higgs mechanism. In this case, the radial component of the Higgs boson is a physical field but it can be decoupled from the dynamics if the radial component is heavy enough. This is achieved by assuming that the gauge coupling constant is much smaller than the Higgs self coupling constant, for example.
Or we can rely on the Stuckelberg mechanism: let the gauge boson mass term be $m^2 g^{\mu\nu}(A_\mu+c\partial_\mu \phi)(A_\nu + c\partial_\nu\phi)$ by introducing additional real scalar field $\phi$. This mass term respects the gauge symmetry $A_\mu\to A_\mu+\partial_\mu\chi$ if $\phi$ transforms as $\phi\to \phi - \chi/c$ with arbitrary function $\chi$. By setting $\phi=0$ using this gauge degree of freedom, we end up with the massive vector boson action (\ref{massive_vector}). In this case, there is no physical degree of freedom other than the vector boson.

It is again convenient to work with the Fourier mode since we are interested in the free vector boson:
\begin{align}
	A_\mu(\vec x,t) = \int \frac{d^3k}{(2\pi)^3} A_{\mu}(\vec k,t) e^{i \vec k\cdot \vec x}.
\end{align}
Since $A_\mu(\vec x,t)$ is a real field, $A_\mu(\vec k,t)=A_\mu^*(-\vec k,t)$ must be satisfied. 
The three physical components are divided into the transverse and longitudinal mode: $\vec A = \vec A_T + \hat k A_L$, where the transverse mode satisfies $\vec k \cdot \vec A_T=0$ with the momentum vector $\vec k$ and $\hat k \equiv \vec k /|\vec k|$.
Note that $A_0$ is not dynamical but it is determined by the constraint equation as
\begin{align}
	A_0(\vec k,t)= \frac{i\vec k \cdot \dot{\vec A}}{k^2+a^2m^2} = \frac{i k\dot{A_L}}{k^2+a^2m^2}  .
\end{align}
Substituting this into the action, we find that the transverse mode and longitudinal mode are decoupled from each other:~\cite{Graham:2015rva}
\begin{align}
	&S = S_T + S_L, \\
	&S_T= \int \frac{d^3k d\tau}{(2\pi)^3} \frac{1}{2}\left( |\partial_\tau \vec A_T|^2 - (k^2+a^2 m^2)|\vec A_T|^2 \right),  \label{ST}\\
	&S_L= \int \frac{d^3k d\tau}{(2\pi)^3} \frac{1}{2}\left( \frac{a^2m^2}{k^2+a^2m^2}|\partial_\tau  A_L|^2 - a^2 m^2|A_L|^2 \right).
\end{align}
The longitudinal mode is further redefined using the canonical field $\widetilde{A_L} \equiv f(\tau) A_L$ with $f(\tau) \equiv am/\sqrt{k^2+a^2m^2}$,
\begin{align}
	S_L= \int \frac{d^3k d\tau}{(2\pi)^3} \frac{1}{2}\left( |\partial_\tau \widetilde{A_L}|^2 - \omega_{L}^2|\widetilde{A_L}|^2 \right),
	~~~~~~\omega_L^2=\frac{a^2m^2}{f^2} -\frac{f''}{f} \equiv k^2 + m_L^2,  \label{SL}
\end{align}
where the effective mass is given by\footnote{
	The same expression was also derived in Ref.~\cite{Ema:2016dny} assuming that the vector boson mass is generated by the Higgs mechanism. 
}
\begin{align}
	m_L^2 = a^2m^2 -\frac{k^2}{k^2+a^2m^2}\left(\frac{a''}{a}-\frac{a'^2}{a^2}\frac{3a^2m^2}{k^2+a^2m^2} \right).
\end{align}

\subsection{Transverse mode production}

From the action of the transverse mode (\ref{ST}) it is evident that it is conformal in the limit $m\to 0$, i.e., the scale factor dependence disappears in the limit $m\to 0$. Hence there is no particle production in this limit. It is similar to the situation of a scalar field with conformal coupling.

To see this in detail, let us expand $\vec A_T$ as
\begin{align}
	\vec A_T(\vec k,\tau) = \sum_{h=\pm}\left[ \mathcal A_T(\vec k, \tau) \vec\epsilon_h a_{\vec k,h} + \mathcal A^*_{T}(\vec k,\tau)\vec{\epsilon_h^*} a_{-\vec k,h}^\dagger \right],
\end{align}
where $\vec \epsilon_h$ denotes the polarization vector for two polarization modes $h=+$ and $-$ which satisfies $\vec{\epsilon_h^*} \cdot \vec \epsilon_{h'} = \delta_{h h'}$. A concrete expression is $\vec\epsilon_{\pm}=(1,\pm i,0)/\sqrt{2}$ if $\vec k$ points to the $z$-direction. The ladder operators satisfy
\begin{align}
	\left[ a_{\vec k,h}, a_{\vec k',h'}^\dagger \right]=(2\pi)^3\delta_{hh'}\delta(\vec k- \vec k'),~~~~~~
	\left[ a_{\vec k,h}, a_{\vec k',h'} \right]=\left[ a_{\vec k,h}^\dagger, a_{\vec k',h'}^\dagger \right]=0.
\end{align}
The mode function obeys the same equation of motion as a scalar field with conformal coupling:
\begin{align}
	\mathcal A_T'' + (k^2 + a^2m^2) \mathcal A_T = 0.   \label{eom_AT}
\end{align}
The solution during inflation (in the de Sitter phase) is given by
\begin{align}
	\mathcal A_T(k,\tau) =e^{\frac{i(2\nu+1)\pi}{4}}\frac{1}{\sqrt{2k}}\sqrt{\frac{-\pi k\tau}{2}} H_\nu^{(1)}(-k\tau),
	~~~~~~\nu^2 \equiv \frac{1}{4}-\frac{m^2}{H^2},
\end{align}
where we have chosen the boundary condition at $k\tau \to-\infty$ (deep inside the horizon) so that it approaches to the mode function in the Minkowski space:
\begin{align}
	\mathcal A_T(k,\tau) \to \frac{1}{\sqrt{2k}}e^{-ik\tau}. \label{BD}
\end{align}
Since $\nu^2 < 1/4$ for $m^2>0$, there is no growth of the superhorizon modes during inflation.
Therefore, the dominant contribution to the gravitational production of the transverse vector boson happens around/after the end of inflation.
We can solve the equation of motion (\ref{eom_AT}) with the initial condition (\ref{BD}) under the background evolution of the cosmic scale factor $a(\tau)$ due to the (spatially homogeneous) inflaton dynamics to determine the energy density of the transverse vector boson through (\ref{rhoT}). 
Instead of solving (\ref{eom_AT}) directly, we can make useful parametrization as follows:
\begin{align}
	\mathcal A_T(k,\tau) = \frac{\alpha_k(\tau)}{\sqrt{2\omega_k}}e^{-i\int^\tau \omega_k(\tau')d\tau'}
	+ \frac{ \beta_k(\tau)  }{\sqrt{2\omega_k}}e^{i\int^\tau \omega_k(\tau')d\tau'},
\end{align}
where $\omega_k \equiv \sqrt{k^2+a^2m^2}$ and $\alpha_k(\tau)$ and $\beta_k(\tau)$ are assumed to satisfy
\begin{align}
	\alpha_k'(\tau) = \frac{\omega_k'}{2\omega_k} e^{2i\int^\tau \omega_k(\tau')d\tau'} \beta_k,~~~~~~
	\beta_k'(\tau) = \frac{\omega_k'}{2\omega_k} e^{-2i\int^\tau \omega_k(\tau')d\tau'} \alpha_k.
\end{align}
It is checked that these set of equations satisfy the equation of motion (\ref{eom_AT}). The initial condition is taken to be $\alpha_k\to 1$ and $\beta_k\to 0$ at $k\tau \to -\infty$. Using this parametrization, the energy density (\ref{rhoT}) is expressed as
\begin{align}
	a^4(\tau) \rho_T(\tau) = 2\int\frac{d^3k}{(2\pi)^3} \omega_k f_T(k,\tau),~~~~~~f_T(k,\tau)=|\beta_k(\tau)|^2.
\end{align}
Here $f_T(k,\tau)$ denotes the occupation number of the transverse vector boson. Note that the normalization condition implies $|\alpha_k(\tau)|^2-|\beta_k(\tau)|^2=1$. In contrast to the case of fermion, this normalization condition does not limit the possible produced number density of vector boson.

Then what we have to do is to estimate $\beta_T(k,\tau)$. The calculation is the same as the gravitational production of a scalar field with conformal coupling as performed in detail in Ref.~\cite{Ema:2018ucl}. Here we present only the results. The number density of the transverse vector boson is given by
\begin{align}
	n_{A_T}(t) \simeq H_{\rm inf}^3\left[ \mathcal C_T \frac{m^4}{m_{\rm inf}^4}+ \eta \frac{m}{H_{\rm inf}} \right] \left( \frac{a(t_{\rm end})}{a(t)} \right)^3,
\end{align}
where $\eta$ is given by the same expression as (\ref{eta}) after reinterpreting $m$ in (\ref{eta}) as the vector boson mass. It is assumed that $m \ll m_{\rm inf}$ since otherwise the vector boson production is suppressed. Taking account of the two polarization degrees of freedom, the numerical coefficient $\mathcal C_T$ is found to be $3/(256\pi)$ if the inflaton potential is well approximated by the quadratic one~\cite{Chung:2018ayg}.
Assuming that the universe is matter-dominated before the completion of reheating, we obtain the energy to entropy density ratio as
\begin{align}
	\frac{\rho_T}{s} = \frac{T_{\rm R} m\,n_{A_T}(t_{\rm end})}{4H_{\rm inf}^2 M_P^2}
	\simeq \frac{T_{\rm R} H_{\rm inf} m}{4M_P^2}
	\left[\mathcal C_T \left(\frac{m}{m_{\rm inf}}\right)^4 + \eta \frac{m}{H_{\rm inf}} \right].
\end{align}

\subsection{Longitudinal mode production}

The longitudinal mode is more similar to a scalar field, as seen from the action (\ref{SL}). It is quantized as
\begin{align}
	\widetilde {A_L}(\vec k,\tau) =\widetilde{\mathcal A_L}(\vec k, \tau)a_{\vec k} +\widetilde{ \mathcal A^*_{L}}(\vec k,\tau) a_{-\vec k}^\dagger,
\end{align}
where the ladder operators satisfy
\begin{align}
	\left[ a_{\vec k}, a_{\vec k'}^\dagger \right]=(2\pi)^3\delta(\vec k- \vec k'),~~~~~~
	\left[ a_{\vec k}, a_{\vec k'} \right]=\left[ a_{\vec k}^\dagger, a_{\vec k'}^\dagger \right]=0.
\end{align}
The equation of motion of the mode function is
\begin{align}
	\widetilde{\mathcal A_L''} + \omega_L^2(k,\tau) \widetilde{\mathcal A_L} =0,
	~~~~~~\omega_L^2 = k^2 + m_L^2.  \label{eom_AL}
\end{align}
For convenience, we also present the equation of motion in the original basis:
\begin{align}
	{\mathcal A_L''} + \frac{2f'}{f} \mathcal A_L' + (k^2+a^2m^2){\mathcal A_L} = 
	{\mathcal A_L''} + \frac{2\mathcal H k^2}{k^2+a^2m^2}\mathcal A_L' + (k^2+a^2m^2){\mathcal A_L} =
	0.
\end{align}

During the de Sitter phase, the effective mass of the longitudinal mode is given by
\begin{align}
	m_L^2 = a^2m^2-a^2 H_{\rm inf}^2 \frac{k^2(2k^2-a^2m^2)}{(k^2+a^2m^2)^2}.
\end{align}
In the high momentum limit $k \gg am$, it is approximated as
\begin{align}
	m_L^2 \simeq a^2(m^2-2H_{\rm inf}^2),
\end{align}
which is the same form as a massive scalar coupled to gravity minimally. Thus we can take the mode function as in the Minkowski form in the high momentum limit $(k/a \gg {\rm max}[m,H_{\rm inf}])$ as
\begin{align}
	\widetilde{\mathcal A_L}(k,\tau) \to \frac{1}{\sqrt{2k}}e^{-ik\tau}.
\end{align}
In the low momentum limit $k \ll am$, we obtain
\begin{align}
	m_L^2 \simeq a^2\left(m^2+H_{\rm inf}^2\frac{k^2}{a^2m^2} \right),
\end{align}
This is always positive definite even in the massless limit $m\to 0$, which is a unique feature of massive vector boson, different from a massive scalar. Now we consider two cases separately: heavy vector boson $m \gtrsim H_{\rm inf}$ and light vector boson $m\lesssim H_{\rm inf}$.

\subsubsection{Heavy vector boson case}

For the heavy vector boson case $m \gtrsim H_{\rm inf}$, it is evident that the effective mass squared $m_L^2$ is always positive independently of the wavenumber $k$, and hence there is no significant growth of the vacuum fluctuation. In particular, no superhorizon modes are enhanced during the de Sitter phase.
Therefore, in this case, we should only take account of the production of high momentum modes after inflation. For $k \gg am$, we obtain
\begin{align}
	\omega_L^2 \simeq k^2 + a^2m^2 - \frac{a''}{a}.
\end{align}
This is the same form as the minimally coupled scalar field. This may be regarded as a consequence of the Goldstone boson equivalence theorem, which says that the longitudinal vector boson may be identified with the Goldstone boson in the high energy limit.
Similar to the case of scalar field, we can again make the following parameterization:
\begin{align}
	\widetilde{\mathcal A_L}(k,\tau) = \frac{\alpha_k(\tau)}{\sqrt{2\omega_L}}e^{-i\int^\tau \omega_L(\tau')d\tau'}
	+ \frac{ \beta_k(\tau)  }{\sqrt{2\omega_L}}e^{i\int^\tau \omega_L(\tau')d\tau'},
\end{align}
where $\alpha_k(\tau)$ and $\beta_k(\tau)$ are assumed to satisfy
\begin{align}
	\alpha_k'(\tau) = \frac{\omega_L'}{2\omega_L} e^{2i\int^\tau \omega_L(\tau')d\tau'} \beta_k,~~~~~~
	\beta_k'(\tau) = \frac{\omega_L'}{2\omega_L} e^{-2i\int^\tau \omega_L(\tau')d\tau'} \alpha_k.
\end{align}
It is again checked that these set of equations satisfy the equation of motion (\ref{eom_AL}). The initial condition is taken to be $\alpha_k\to 1$ and $\beta_k\to 0$ at $k\tau \to -\infty$. The energy density (\ref{rhoL}) is expressed as
\begin{align}
	a^4(\tau) \rho_L(\tau) = 2\int\frac{d^3k}{(2\pi)^3} \omega_L f_L(k,\tau),~~~~~~f_L(k,\tau)=|\beta_k(\tau)|^2,
\end{align}
where $f_L(k,\tau)$ denotes the occupation number of the longitudinal vector boson. The normalization condition implies $|\alpha_k(\tau)|^2-|\beta_k(\tau)|^2=1$.
In this case, therefore, the number density of produced longitudinal vector boson is estimated in the same as a minimal scalar field~\cite{Ema:2018ucl},
\begin{align}
	n_{A_L}(t) \simeq \mathcal C_L H_{\rm inf}^3\left( \frac{a(t_{\rm end})}{a(t)} \right)^3,
\end{align}
where $\mathcal C_L \sim 3/(512\pi)$~\cite{Chung:2018ayg} and we assumed $m < m_{\rm inf}$. Compared with the transverse mode, there is no suppression factor of $(m/m_{\rm inf})^4$. Thus the energy to entropy density ratio is evaluated as
\begin{align}
	\frac{\rho_L}{s} = \frac{T_{\rm R} m\,n_{A_L}(t_{\rm end})}{4H_{\rm inf}^2 M_P^2}
	\simeq \frac{\mathcal C_L T_{\rm R} H_{\rm inf} m}{4M_P^2}.
\end{align}

\subsubsection{Light vector boson case}

For the light vector boson case $m \lesssim H_{\rm inf}$, the situation is different.
Superhorizon modes with $m < k/a < H_{\rm inf}$ experience tachyonic growth during inflation, similar to the case of light scalar and these inflation-generated long wavelength modes may give dominant contributions to the final vector boson abundance.
Note that the growth is eventually terminated when the physical wavenumber $k/a$ becomes equal to the vector boson mass $m$. Thus the power spectrum of such a massive vector boson at large scale is much suppressed compared with the case of light scalar field with the same mass~\cite{Graham:2015rva}.
For modes with $m < k/a < H_{\rm inf}$, it is easily found from (\ref{eom_AL}) that the mode function grows as\footnote{
	Note that the original field $A_L(k,\tau)$ before the canonical rescaling remains constant.
}
\begin{align}
	\widetilde{\mathcal A_L}(k,\tau) \propto \tau^{-1} \propto a.
\end{align}
Therefore, at the end of inflation, the mode function becomes
\begin{align}
	\left| \widetilde{\mathcal A_L}(k) \right|^2 \simeq \begin{cases}
		\displaystyle \frac{1}{2k}  &{\rm for}~~ k > a_{\rm end} H_{\rm inf}\\
		\displaystyle \frac{1}{2k}\left( \frac{a_{\rm end} H_{\rm inf}}{k} \right)^2 &{\rm for}~~ a_{\rm end}m < k < a_{\rm end} H_{\rm inf} \\
		\displaystyle \frac{1}{2k}\left( \frac{H_{\rm inf}}{m} \right)^2 & {\rm for}~~ k< a_{\rm end} m
	\end{cases}.
\end{align}
In terms of the energy density per log frequency at the end of inflation, after the renormalization of the UV divergence as usual, we have
\begin{align}
	\rho_L(k,t_{\rm end}) \simeq 
		 \frac{1}{2}\left(\frac{k}{a_{\rm end}} \right)^2 \left(\frac{H_{\rm inf}}{2\pi}\right)^2 ~~{\rm for}~~  k < a_{\rm end} H_{\rm inf}.
\end{align}
At this stage, therefore, the shortest wavelength mode $(k\sim H_{\rm inf}a_{\rm end})$ gives the dominant contribution to the total energy density of the longitudinal vector boson. However, these modes are (highly) relativistic and receive larger suppression factor due to the redshift.
As shown in Appendix, the energy density scales as 
\begin{align}
	\rho_L(k) \propto \begin{cases}
		a^{-2} &{\rm for}~~  k< aH~~{\rm and}~~m<H \\
		a^{-3} & {\rm for}~~ k< am~~{\rm and}~~m>H  \\
		a^{-4} & {\rm for}~~ k > am~~{\rm and}~~k>aH
	\end{cases}.
\end{align}
Assuming again that the universe is matter-dominated before the completion of the reheating,
the final energy to entropy density ratio is expressed in terms of the number density at $H=m$, defined by $n_L(k;H=m) = \rho_L(k;H=m)/\sqrt{m^2+k^2/a^2(H=m)}$:
\begin{align}
	\frac{\rho_L(k)}{s} \simeq \begin{cases}
		\displaystyle \frac{T_{\rm R} n_L(H=m)}{4m M_P^2} &{\rm for}~~ H_{\rm R} < m \\
		\displaystyle \frac{T_{\rm R} n_L(H=m)}{4m M_P^2}\left(\frac{m}{H_{\rm R}}\right)^{1/2}
		=\left(\frac{90}{\pi^2 g_*}\right)^{1/4} \frac{n_L(H=m)}{4m^{1/2} M_P^{3/2}} &{\rm for}~~ H_{\rm R} > m 
	\end{cases},
\end{align}
where $H_{\rm R} \equiv \sqrt{\pi^2g_*/90}\,T_{\rm R}^2/M_P$ is the Hubble parameter at the completion of reheating.

For $H_{\rm R} < m$, we find
\begin{align}
	\frac{\rho_L(k)}{s} \simeq \begin{cases}
		\displaystyle \frac{T_{\rm R} H_{\rm inf}^{2/3}}{32\pi^2 m^{2/3} M_P^2}\left( \frac{k}{a_{\rm end}} \right)^2 &{\rm for}~~ k< k_* \\
		\displaystyle \frac{T_{\rm R} mH_{\rm inf}^4}{32\pi^2M_P^2}\left( \frac{a_{\rm end}}{k} \right)^3 &{\rm for}~~ k> k_* 
	\end{cases}~~~(H_{\rm R} < m),
\end{align}
where $k_*\equiv a(H=m) m$ denotes the comoving wavenumber for which the corresponding mode becomes non-relativistic at $H=m$. Thus it is seen that the total energy density is dominated by modes around $k=k_*$. In this case we have $k_* = a_{\rm end}m (H_{\rm inf}/m)^{2/3}$. The total vector boson abundance is then given by
\begin{align}
	\frac{\rho_L}{s} \simeq \frac{T_{\rm R} H_{\rm inf}^2}{32\pi^2 M_P^2}
	\simeq 5\times 10^{-10}\,{\rm GeV}\left(\frac{T_{\rm R}}{10^6\,{\rm GeV}}\right)\left(\frac{H_{\rm inf}}{10^{12}\,{\rm GeV}}\right)^2.
\end{align}
It is independent of the vector boson mass $m$.

For $H_{\rm R} > m$, after straightforward but tedious calculations, we find
\begin{align}
	\frac{\rho_L(k)}{s} \simeq \begin{cases}
	\displaystyle \frac{T_{\rm R}H_{\rm inf}^{2/3}}{32\pi^2 M_P^2 m^{1/2}H_{\rm R}^{1/6}}\left( \frac{k}{a_{\rm end}} \right)^2 &{\rm for}~~k < k_* \\
		\displaystyle \frac{T_{\rm R} m H_{\rm inf}^{8/3}}{32\pi^2 M_P^2 H_{\rm R}^{2/3}}\left( \frac{a_{\rm end}}{k} \right) &{\rm for}~~k_* < k < a(H=H_{\rm R}) H_{\rm R} \\
		\displaystyle \frac{T_{\rm R} mH_{\rm inf} ^4}{32\pi^2M_P^2}\left( \frac{a_{\rm end}}{k} \right)^3 &{\rm for}~~ k> a(H=H_{\rm R}) H_{\rm R} 
	\end{cases}~~~(H_{\rm R} > m).
\end{align}
Again we find that it is peaked around $k= k_*$ where $k_* \equiv a(H=m)m$, which is now evaluated as $k_*= a_{\rm end} m(H_{\rm inf}/H_{\rm R})^{2/3} (H_{\rm R}/m)^{1/2}$. Thus the total vector boson abundance is given by
\begin{align}
	\frac{\rho_L}{s} \simeq \left(\frac{90}{\pi^2 g_*}\right)^{1/4} \frac{m^{1/2} H_{\rm inf}^2}{32\pi^2 M_P^{3/2}}
	\simeq 5\times 10^{-10}\,{\rm GeV}\left(\frac{m}{10^{-6}\,{\rm GeV}}\right)^{1/2}\left(\frac{H_{\rm inf}}{10^{12}\,{\rm GeV}}\right)^2.
\end{align}
It is consistent with Ref.~\cite{Graham:2015rva}.


\subsection{Combined Results}

Let us summarize the results so far. The abundance of longitudinal vector boson which is purely gravitationally produced is given by
\begin{align}
	\frac{\rho_L}{s} \simeq \begin{cases}
		\displaystyle \frac{\mathcal C_L T_{\rm R} H_{\rm inf} m}{4M_P^2} & {\rm for}~~ H_{\rm inf} <  m \\
		\displaystyle \frac{T_{\rm R} H_{\rm inf}^2}{32\pi^2 M_P^2}& {\rm for}~~ H_{\rm R} < m < H_{\rm inf} \\
		\displaystyle \left(\frac{90}{\pi^2 g_*}\right)^{1/4} \frac{m^{1/2} H_{\rm inf}^2}{32\pi^2 M_P^{3/2}}& {\rm for}~~m< H_{\rm R}
	\end{cases}.
\end{align}
Note that the origin of the vector boson is different between the case of $m> H_{\rm inf}$ and $m< H_{\rm inf}$. In the former case, the vector boson is dominantly produced at the end of inflation or during the early stage of reheating and the main produced mode is about the inflaton mass: $k\sim a_{\rm end} m_{\rm inf}$. In the latter case, the dominant contribution comes from the superhorizon mode generated during inflation, which eventually re-enters the horizon at $H \sim m$.
The transverse modes are also produced at the end of inflation and during the reheating stage, but they are always subdominant compared with the longitudinal mode.

\begin{figure}[t]
	\begin{center}
		\includegraphics[scale=0.35]{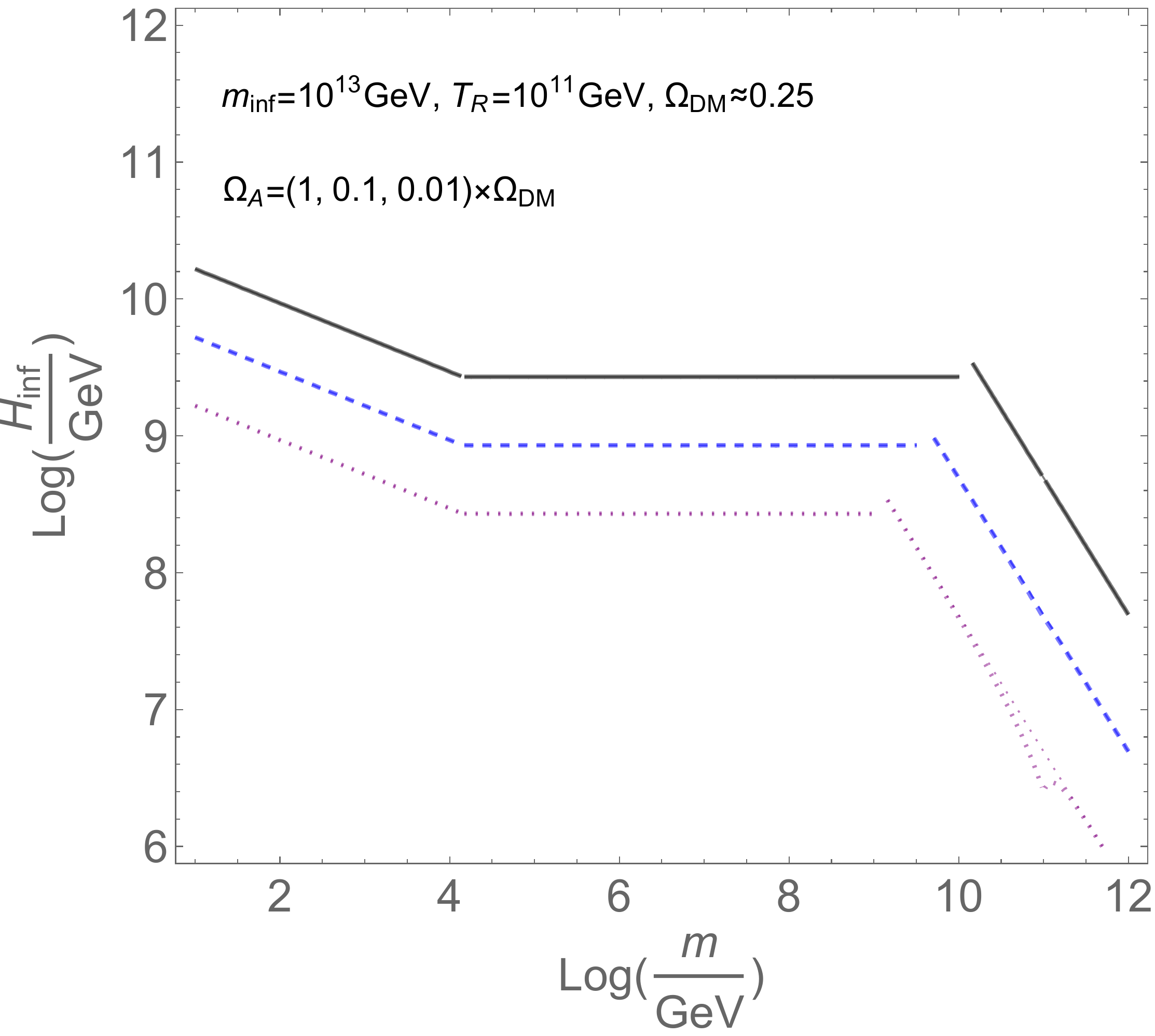}
		\includegraphics[scale=0.35]{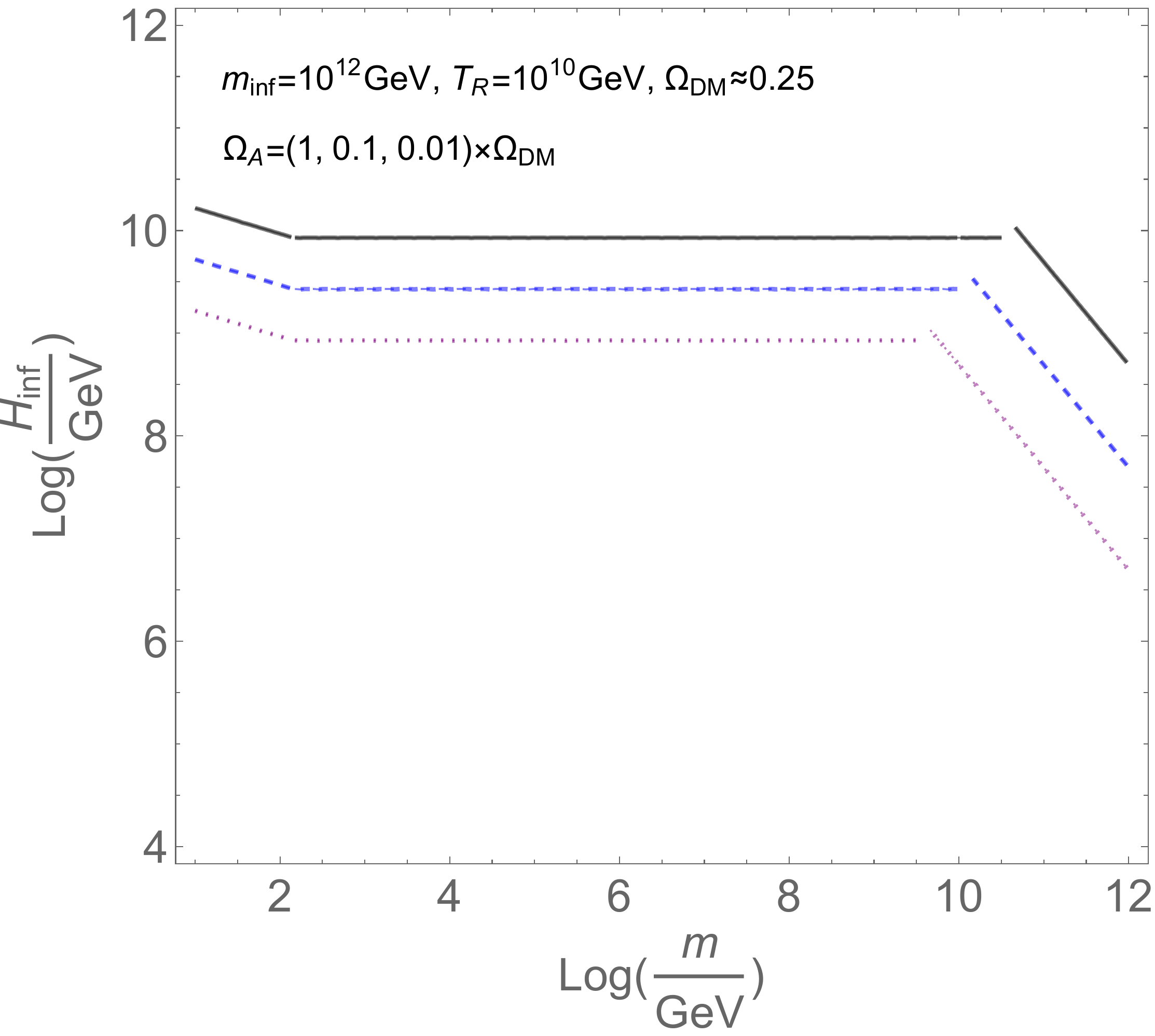}
	\end{center}
	\caption{Illustration of the gravitationally produced vector boson abundance with two sets of inflaton mass and  reheating temperature, $m_{\textrm{inf}}=10^{13}\,$GeV and $T_{\rm R}=10^{11}\,$GeV (Left), $m_{\textrm{inf}}=10^{12}\,$GeV and $T_{\rm R}=10^{10}\,$GeV (Right). 
		Three different curves (gray solid, blue dashed, and purple dotted) correspond to $\Omega_{A}=(1,0.1,0.01)\times\Omega_{\textrm{DM}}$. 
		\label{fig:vectorL}	}
\end{figure}

In Fig.~\ref{fig:vectorL}, we show several contours of the vector boson abundance on the parameter space of $(m,H_{\rm inf})$ for two sets of inflaton mass and  reheating temperature, $m_{\textrm{inf}}=10^{13}\,$GeV and $T_{\rm R}=10^{11}\,$GeV (Left panel), $m_{\textrm{inf}}=10^{12}\,$GeV and $T_{\rm R}=10^{10}\,$GeV (Right panel). Similar to the fermion case, thermal production is included, see Appendix~\ref{sec:thermalp}. Three contours (gray solid, blue dashed, and purple dotted) correspond to $\Omega_{A}=(1,0.1,0.01)\times\Omega_{\textrm{DM}}$ where $\Omega_A = \rho_A/ \rho_{\rm crit}$ is the density parameter of the vector boson. We can see there are wide and viable parameter regions that can satisfy the current DM relic abundance. Contours without including thermal productions are shown in thin curves that however almost coincide with thick ones, which means thermal contributions are negligible in the showed parameter space. 

\section{Conclusions and discussion}  \label{sec:conc}

We have studied the DM production mechanism in the case where the DM particle is a massive fermion or vector boson and has only the gravitational interaction. The production takes place through the so-called gravitational particle production under the standard inflationary cosmology.  

For the case of a massive fermion, the presence of mass term violates the conformal invariance and it somehow feels the background time evolution, resulting in particle production. The dominant production process depends on the fermion's mass $m$. For $m\lesssim H_{\rm inf}$, the non-adiabaticity of the fermion wave function is prominent when the fermion becomes non-relativistic $k\sim am$ for each wavenumber $k$. Those with momentum $k$ such that $k\sim am$ and $H \sim m$ gives the dominant contribution to the final fermion abundance as already pointed out in Ref.~\cite{Chung:2011ck}. For $m\gg H_{\rm inf}$, such an effect of the universe expansion is negligible while the inflaton coherent oscillation produces excites the high momentum fermion modes, since the cosmic scale factor $a(\tau)$ includes a small but nonzero oscillating part. In both cases, we have the viable parameter regions that can reproduce the present DM abundance. All these features are similar to the case of a scalar field with conformally coupled to gravity~\cite{Ema:2018ucl}.

For the case of a massive vector boson, the story is a bit complicated. The transverse mode is conformal in the massless limit, and hence the gravitational production proceeds only through the presence of mass term. Again it is similar to the case of conformally coupled scalar field. On the other hand, the longitudinal mode shows more non-trivial behavior. For $m\lesssim H_{\rm inf}$, during the de Sitter phase the vector obtains superhorizon quantum fluctuations and eventually behaves as non-relativistic matter. In contrast to the scalar field with minimal coupling, there is a limit for the growth of the superhorizon modes at $k\sim am$, and hence such a model is not constrained by the presence of DM isocurvature perturbation on cosmological scales~\cite{Graham:2015rva}. For $m\gg H_{\rm inf}$, it is rather close to the minimally coupled scalar field, and the inflaton coherent oscillation excites the high-momentum longitudinal mode. In both cases we have correct parameter regions that reproduce the present DM abundance.

Several comments are in order. For analyses of both the fermion and vector boson, we have assumed that the fermion/vector boson mass parameter $m$ is just a constant. It is possible that the mass is given by the expectation of value of some other Higgs-like scalar field. In such a case the mass parameter $m$ can be dynamical. For example, it can have a different value in the inflationary era and the present universe. Our results crucially depend on this assumption. If there is a Higgs field, we should take account of the dynamics of the Higgs field if the mass of the Higgs field is smaller than the Hubble parameter $H$, which may significantly affect our estimate of the fermion and vector boson abundance.

Finally we comment on the detectability of our model. Since DM is only gravitationally interacting, it is hopeless to find it through a kind of direct detection experiments. If DM is long-lived but has finite lifetime, its decay would produce high-energy cosmic rays for indirect detections~\cite{Ibarra:2013cra}. 
As seen from Figs.~\ref{fig:fermion} and \ref{fig:vectorL}, the wide DM mass range is consistent with the DM abundance and correspondingly the energy of cosmic-rays induced by the DM decay also can take wide range of values. Another interesting possibility may be to search for the effect of heavy field (say, $m\sim H_{\rm inf}$) through the non-Gaussianity of the primordial curvature fluctuations~\cite{Chen:2009zp,Baumann:2011nk,Assassi:2012zq,Noumi:2012vr,Arkani-Hamed:2015bza,Li:2019ves}.

\section*{Acknowledgments}

The work of Y.E. was supported in part by JSPS Research Fellowships for Young Scientists.
This work was supported by the Grant-in-Aid for Scientific Research C (No.18K03609 [KN]) and Innovative Areas (No.26104009 [KN], No.15H05888 [KN], No.17H06359 [KN], 16H06490 [YT]).

\appendix

\section{Conventions}

Our convention follows Ref.~\cite{Freedman:2012zz}. The flat space metric is taken to be $\eta_{\mu\nu}={\rm diag}(-1,1,1,1)$. The gamma matrices in the flat space is denoted by $\gamma_a$ while those in the curved background are expressed by $\hat\gamma_\mu$. They are related by $\hat\gamma_\mu=e^a_\mu \gamma_a$ by using the vierbein $e^a_\mu$. In the FRW background, the vierbein is 
\begin{align}
	e^{a}_\mu = a \delta^a_\mu,~~~~~e^\mu_a=\frac{1}{a}\delta^\mu_a,
\end{align}
so that the metric is given by $g_{\mu\nu}=e^a_\mu e^b_\nu \eta_{ab} = a^2 \eta_{\mu\nu}$. 
The Clifford algebra is then defined as
\begin{align}
	\left\{ \hat\gamma^\mu, \hat\gamma^\nu \right\} = 2g^{\mu\nu}.
\end{align}
Thus we have $(\hat \gamma^0)^\dagger = -\hat\gamma^0$ and $(\hat \gamma^i)^\dagger = \hat \gamma^i$. We also define $\gamma_5 \equiv i \gamma_0 \gamma_1 \gamma_2 \gamma_3$.
The explicit Weyl representation of the gamma matrices is
\begin{align}
	\gamma_\mu = \begin{pmatrix}
		0 & \sigma^\mu \\
		\bar\sigma^\mu & 0
	\end{pmatrix},~~~~~~
	\gamma_5 = \begin{pmatrix}
		1 & 0 \\
		0 & -1
	\end{pmatrix},
\end{align}
where $\sigma_\mu=(-1,\sigma_i)$ and $\bar\sigma_\mu=(1,\sigma_i)$ with $\sigma_i$ being the usual Pauli matrices that satisfy $\{\sigma_i,\sigma_j\}=2\delta_{ij}$. In the Dirac representation the explicit form of the gamma matrices is
\begin{align}
	\gamma_0 = \begin{pmatrix}
		i & 0 \\
		0 & -i
	\end{pmatrix},~~~~~~
	\vec\gamma = \begin{pmatrix}
		0 & i\vec\sigma \\
		-i\vec\sigma & 0
	\end{pmatrix},~~~~~~
	\gamma_5 = \begin{pmatrix}
		0 & 1\\
		1 & 0
	\end{pmatrix}.
\end{align}

\section{Energy density of vector boson}

\subsection{Energy-momentum tensor and energy density}

The energy momentum tensor is defined as
\begin{align}
	T_{\mu\nu} = \frac{-2}{\sqrt{-g}}\frac{\delta(\sqrt{-g}\mathcal L)}{\delta g^{\mu\nu}}.
\end{align}
For a massive vector boson, it is given by
\begin{align}
	T_{\mu\nu}=g^{\rho\sigma}F_{\mu\rho}F_{\nu\sigma}-\frac{1}{4}g_{\mu\nu}g^{\rho\alpha}g^{\sigma\beta}F_{\rho\sigma}F_{\alpha\beta}
	+ m^2\left(A_\mu A_\nu-\frac{1}{2}g_{\mu\nu}g^{\rho\sigma}A_\rho A_\sigma\right).
\end{align}

After some calculations, we find that the energy density $\rho= \left< 0| T_{00} |0\right>$ is given by
\begin{align}
	&\rho= \rho_T + \rho_L, \\
	&\rho_T = 2\int \frac{d^3k}{(2\pi)^3} \frac{1}{2a^4}\left[ | {\mathcal A_T'}(k)|^2 + (k^2+a^2m^2) |{\mathcal A_T}(k)|^2 \right],\label{rhoT}\\
	&\rho_L = \int \frac{d^3k}{(2\pi)^3} \frac{1}{2a^4}\left[ \frac{a^2m^2}{k^2+a^2m^2}|\mathcal A_L'(k)|^2 + a^2m^2 |\mathcal A_L(k)|^2 \right], \label{rhoL}
\end{align}
where the prefactor $2$ in the expression of $\rho_T$ accounts for the two polarization states. Again we can rewrite the longitudinal mode by using the canonical field $\widetilde A_L(\vec k)=f A_L(\vec k)$ with $f(\tau) \equiv am/\sqrt{k^2+a^2m^2}$,
\begin{align}
	\rho_L &= \int \frac{d^3k}{(2\pi)^3} \frac{1}{2a^4}\left[ |\widetilde {\mathcal A_L'}(k)|^2 + \frac{a^2m^2+f'^2}{f^2}|\widetilde {\mathcal A_L}(k)|^2 -\frac{f'}{f}
	\left(\widetilde {\mathcal A_L'}(k)\widetilde {\mathcal A_L^*}(k)+\widetilde {\mathcal A_L'^*}(k)\widetilde {\mathcal A_L}(k) \right) \right], \nonumber\\
	&=\int \frac{d^3k}{(2\pi)^3} \frac{1}{2a^4}\left[ |\widetilde {\mathcal A_L'}(k)|^2 + \left(k^2+a^2m^2 + \mathcal H^2\left(\frac{k^2}{k^2+a^2m^2}\right)^2 \right)|\widetilde {\mathcal A_L}(k)|^2 \right. \nonumber \\
	&~~~~~~~~~~~~~\left. -\mathcal H \frac{k^2}{k^2+a^2m^2}\left(\widetilde {\mathcal A_L'}(k)\widetilde{\mathcal  A_L^*}(k)+\widetilde {\mathcal A_L'^*}(k)\widetilde{\mathcal  A_L}(k) \right) \right].
\end{align}
Note that these expressions show UV divergence when the vacuum configuration for the mode function is substituted, which must be subtracted to obtain nearly vanishing cosmological constant observed now.

\subsection{Energy scaling of the vector boson}

In this appendix we present approximate solutions to the equation of motion of the vector boson mode function after inflation and show how the corresponding energy density scales with respect to the cosmic scale factor $a(\tau)$. Hereafter we assume the reheating phase of the universe with equation of state $w$ such that $H\propto a^{-3(1+w)/2}$.

\subsubsection{Transverse mode}

The equation of motion for the transverse mode is 
\begin{align}
	\mathcal A_T'' + (k^2 + a^2m^2) \mathcal A_T = 0.
\end{align}

In the light vector boson regime $(m \ll H)$, the solution looks like\footnote{
	Actually there is also a solution $\mathcal A_T \propto a^{(1+3w)/2}$ for $k\ll am$ and this is a growing solution for $w> -1/3$.  It is not trivial whether this growing solution is of physical importance or not, but detailed investigations show that the boundary condition (\ref{BD}) does not lead to this growing solution~\cite{Graham:2015rva}.
}
\begin{align}
	\mathcal A_T \sim \begin{cases}
		\displaystyle \frac{1}{\sqrt{2k}} e^{ik\tau} & {\rm for}~~k \gg am \\
		\displaystyle \frac{1}{\sqrt{2k}} & {\rm for}~~k \ll am
	\end{cases}.
\end{align}
Thus $\rho_T \propto a^{-4}$ for $k\gg am$ and $\rho_T \propto a^{-2}$ for $k\ll am$.

In the heavy vector boson regime $(m \gg H)$, the solution looks like
\begin{align}
	\mathcal A_T \sim \begin{cases}
		\displaystyle \frac{1}{\sqrt{2k}} e^{ik\tau} & {\rm for}~~k \gg am \\
		\displaystyle \frac{1}{\sqrt{2am}} e^{im\int ad\tau} & {\rm for}~~k \ll am
	\end{cases},
\end{align}
Thus $\rho_T \propto a^{-4}$ for $k\gg am$ and $\rho_T \propto a^{-3}$ for $k\ll am$, as just expected from the scaling of relativistic and non-relativistic matter, respectively.

\subsubsection{Longitudinal mode}

The equation of motion for the longitudinal mode is 
\begin{align}
	\widetilde{\mathcal A_L}'' + \omega_L^2 \widetilde{\mathcal A_L} = 0,~~~~~\omega_L^2=k^2+m_L^2,
\end{align}
where the effective mass is given by\footnote{
	Some relations in the universe with equation of state $w$ are: $a''/a=\mathcal H^2(1-3w)/2$, $\mathcal H \tau =2/(1+3w)$,
	 $\mathcal H' =-\mathcal H^2(1+3w)/2$, $a\propto \tau^{2/(1+3w)}$
}
\begin{align}
	m_L^2= a^2m^2 - \frac{k^2}{k^2+a^2m^2}\left(\frac{1-3w}{2}-\frac{3a^2m^2}{k^2+a^2m^2}\right) \mathcal H^2.
\end{align}

In the case of light vector boson regime $(m\ll H)$, it is a bit complicated:
\begin{align}
	\omega_L^2 \simeq \begin{cases}
		\displaystyle k^2- \left(\frac{1-3w}{2} -\frac{3a^2m^2}{k^2}\right)\mathcal H^2 & {\rm for}~~k \gg am \\
		\displaystyle a^2m^2 + \frac{5+3w}{2} \frac{k^2}{a^2m^2}\mathcal H^2 & {\rm for}~~k \ll am
	\end{cases}.
	\label{omegaL_light}
\end{align}
Thus the solution looks like
\begin{align}
	\widetilde{\mathcal A_L}\propto \begin{cases}
		\displaystyle e^{ik\tau} & {\rm for}~~aH\ll k\\
		\displaystyle c_1 a + c_2 a^{(3w-1)/2} & {\rm for}~~am\ll k \ll aH \\
		\displaystyle c_3 + c_4 a^{(3w+1)/2} & {\rm for}~~k \ll am
	\end{cases}.
\end{align}
with $c_1$--$c_4$ denoting numerical constants determined by the initial conditions at the end of inflation.\footnote{
	These solutions can be found by substituting the ansatz $\widetilde{\mathcal A_L} \propto a^\ell$ and requiring that leading terms vanish to determine the exponent $\ell$. Here we neglect the last terms in (\ref{omegaL_light}) since they do not much affect the value of $\ell$.
}
Note again that the growing solution for $k\ll am$ actually is not of physical importance if one properly connects the solution during inflation to that during the reheating phase~\cite{Graham:2015rva}. Thus we only need to see terms with $c_1$ and $c_3$ for $k\ll aH$. As a result, we obtain the scaling of the longitudinal vector boson energy density as $\rho_L \propto a^{-4}$ for $k \gg aH$ (subhorizon modes) and $\rho_L \propto a^{-2}$ for $k \ll aH$ (superhorizon modes).

In the case of heavy vector boson regime $(m\gg H)$, on the other hand, we simply obtain
\begin{align}
	\omega_L^2 \simeq k^2 + a^2m^2.
\end{align}
It is the same as the transverse one and hence we obtain $\rho_L \propto a^{-4}$ for $k\gg am$ and $\rho_L \propto a^{-3}$ for $k\ll am$, just as expected.

\section{Thermal production}\label{sec:thermalp}

Here we briefly summarize the thermal production (TP) of gravitationally interacting DM.
The production cross section of purely gravitational DM $X$, either $X$ is fermion or vector boson, through the scattering of SM particles in the thermal bath with the temperature $T$ is~\cite{Garny:2015sjg,Tang:2016vch,Tang:2017hvq,Garny:2017kha}
\begin{align}
\left< \sigma v\right> = y\frac{T^2}{M_P^4}~~~~~~{\rm for}~~~T > m_X,
\end{align}
with some numerical constant $y, y\simeq 0.2$ for fermions and $y\simeq 0.8$ for vectors. See Ref.~\cite{Tang:2017hvq} for details. The DM number density created per Hubble time is given by
\begin{align}
n_X \sim n_{\rm SM}^2 \left< \sigma v\right> H^{-1} \sim 
\begin{cases}
\displaystyle  y\frac{T_{\rm R}^4 H}{M_P^2} \propto a^{-3/2} & {\rm for}~~~T > T_{\rm R} \\
\displaystyle y\frac{T^6}{M_P^3}  \propto a^{-6} & {\rm for}~~~T < T_{\rm R}
\end{cases},
\end{align}
assuming $T > m_X$. Otherwise, the production rate is suppressed by $e^{-m_X/T}$.
Therefore, if $T_{\rm R} > m_X$, the production is dominated at $T \sim T_{\rm R}$.
On the other hand, if $T_{\rm R} < m_X$, the production is dominated at $T \sim m_X$.
Thus the DM abundance through the gravitational annihilation of SM particles is given by
\begin{align}
\left(\frac{\rho_X}{s}\right)^{\rm (TP)} \simeq 
\frac{m_X n_X(T_{\rm prod})}{3 H_{\rm prod}^2 M_P^2} \frac{3 T_{\rm R}}{4}.
\end{align}
Here the subscript ``prod'' refers to the dominant production epoch. It is estimated as
\begin{align}
\left(\frac{\rho_X}{s}\right)^{\rm (TP)} \sim
\begin{cases}
\displaystyle \tilde y\frac{T_{\rm R}^7}{m_X^3M_P^3}
\sim 10^{-15}\,{\rm GeV}\,\tilde y \left( \frac{T_{\rm R}}{10^{10}\,{\rm GeV}} \right)^7
\left( \frac{10^{10}\,{\rm GeV}}{m_X} \right)^3 
& {\rm for}~m_X > T_{\rm R} \\
\displaystyle \tilde y\frac{m_XT_{\rm R}^3}{M_P^3} 
\sim 10^{-15}\,{\rm GeV}\,\tilde y \left( \frac{T_{\rm R}}{10^{10}\,{\rm GeV}} \right)^3
\left( \frac{m_X}{10^{10}\,{\rm GeV}} \right)
& {\rm for}~m_X < T_{\rm R}
\end{cases},
\label{rhosTP}
\end{align}
where $\tilde y$ includes various $\mathcal O(1)$ numerical factor neglected in the rough estimation.
This should be compared with the gravitational production studied in the main text.
For the case of vector boson or minimal scalar, it can dominate over the gravitational production if the reheating is close to instantaneous, i.e., if $H_{\rm R}$ is not very far from $H_{\rm inf}$. For the fermion case, it depends on the model parameters $m_X, H_{\rm inf}, T_{\rm R}$ and $m_{\rm inf}$ in a more complicated way.



\end{document}